\documentclass[pra,twocolumn,nofootinbib,eqsecnum,floatfix,showpacs]{revtex4}
\usepackage{bm} % bold math
\usepackage{amsmath}

\begin{document}

\title{{\bf Typicality Derived}
\thanks{Alberta-Thy-01-08, arXiv:0804.3592 [hep-th]}}

\author{Don N. Page}
\email{don@phys.ualberta.ca}

\affiliation{Institute for Theoretical Physics\\
Department of Physics, University of Alberta\\
Room 238 CEB, 11322 -- 89 Avenue\\
Edmonton, Alberta, Canada T6G 2G7\footnote{Permanent address},}
\affiliation{Asia Pacific Center for Theoretical Physics, Pohang
790-784, Korea}

\date{2008 June 3}

\begin{abstract}

Hartle and Srednicki have suggested that standard quantum theory does not
favor our typicality.  Here an alternative version is proposed in which
typicality is likely, {\it Eventual Quantum Mechanics\/}.  This version
allows one to calculate normalized probabilities for alternatives obeying
what I call the {\it Principle of Observational Discrimination\/}, that
each possible complete observation or data set should uniquely distinguish
one element from the set of alternatives.

\end{abstract}


\maketitle

\section{Introduction}

Hartle and Srednicki \cite{HS} use a type of probabilistic reasoning that
includes standard quantum theory to argue that ``it is perfectly possible
(and not necessarily unlikely) for us to live in a universe in which we are
not typical.''  However, this leads to their conclusion (iv): 
``Cosmological models that predict that at least one instance of our data
exists (with probability one) somewhere in spacetime are indistinguishable
no matter how many other exact copies of these data exist.''  If one were
forced to abide by that limitation, then \cite{DP07typ} a huge variety of
cosmological models giving sufficiently large universes would predict
nearly unit probability for our data set and hence the same likelihoods. 
Thus observations would count for nothing in distinguishing between these
theories, and much of cosmology would cease to be an observational science.

Hartle and Srednicki \cite{HS} note that a common kind of reasoning in
cosmology starts from an assumption that some property of human observers
is typical.  They cite \cite{BLL94,Vil95,DKS02,Page06a,BF06,Page06b,
Linde06,Page06c,Vil06,Page06d,Banks07} as giving examples of this
reasoning, which they question.  They point out that this reasoning would
not be valid in a version of quantum theory in which highly atypical
observations are not highly unlikely.  On the other hand, more recent
arguments against the conclusions of Hartle and Srednicki have been given
in \cite{DP07typ,DP07nobang,GV07,DP07obs,BFY07}.  For example, Bousso,
Freivogel, and Yang argue \cite{BFY07} that ``the Hartle-Srednicki
prescription would put an end to experimental science.  It would render all
experiments pointless, because we could not reject any theory until we know
how many other laboratories there are.  Given the success of the scientific
method thus far, we may conclude the Hartle-Srednicki prescription is
inappropriate.''

How can we rescue science from the dire conclusions Hartle and Srednicki
draw from standard quantum theory and other similar types of probabilistic
reasoning?  Here I argue that this can be done by reformulating quantum
theory so that it gives normalizable probabilities for the alternatives of
all possible distinct observations.
%rather than for more restricted classes.

Here I shall define {\it standard quantum theory} to be any version of
quantum theory in which observably distinct alternatives are restricted
to orthogonal projection operators (with the probabilities of these
alternatives being given by the expectation values of the corresponding
orthogonal projection operators).  Such a quantum theory may be suitable
for quantum states in which there are no more than one copy of any observer
(or set of communicating observers, or civilization, or human scientific
information gathering and utilizing system, HSI \cite{HS}, though here I
shall henceforth just say ``observer'' for any of these possibilities). 
Then different possible observations by that observer presumably can be
described by orthogonal projection operators.  However, for cosmological
quantum states for a universe sufficiently large that there is more than
one copy of an observer that can jointly make distinct observations, these
distinct observations need not correspond to orthogonal projection
operators.  Therefore, standard quantum theory is not able to assign
normalizable probabilities to such sets of distinct alternative
observations.

For example, suppose we consider the observation of how many heads occur
when two coins are tossed in a certain recorded way.  There are three
possible distinct observations for the numbers of heads that occur in one
tossing of two coins (0, 1, and 2).  If only one set of two coins is tossed
(e.g., by only one observer), then these distinct observations can be
assigned orthogonal projection operators.  If one has a quantum state in
which it is definitely true that exactly one set of two coins is tossed,
and each head is observed to land definitely heads or tails, then the
expectation value of each of the three projection operators is interpreted
in standard quantum theory to be the probability for that number of heads,
and these probabilities are normalized to sum to unity.

However, if there is more than one tossing of two coins each (say by more
than one copy of an observer), then distinct observations for the numbers
of heads do {\it not\/} correspond to orthogonal projection operators.  For
example, one can have $N_1$ heads in one of the tossings (say by one copy
of the observer) and $N_2$ heads in a second tossing of two coins (say by
another copy of the observer), and even if $N_1\neq N_2$, these distinct
observations do not correspond to orthogonal projection operators.  In
nonquantum language, one says that these two distinct observations are not
mutually exclusive, since both can occur (one for each copy of the
observer).  If one calculates the expectation values of the projection
operators corresponding to all the distinct three observations of the
number of heads in a tossing, these expectation values will have a sum that
is greater than unity.

For example, if the coin is fair, then for one tossing of two coins the
probability of 0 heads is 1/4, of 1 head is 1/2, and of 2 heads is 1/4,
which sum to unity.  However, for two tossings of two coins each, the
probability is $1-(1-1/4)^2 = 7/16$ for the existence of an observation of
0 heads, $1-(1-1/2)^2 = 3/4$ for the existence of an observation of 1 head,
and $1-(1-1/4)^2 = 7/16$ for the existence of an observation of 0 heads,
which sum to 13/8, greater than unity.  That is, the expectation value for
the projection operator for at least one observation of 0 heads is 7/16,
for at least one observation of 1 head is 3/4, and for at least one
observation of 2 heads is 7/16.  When one has in mind a view that
encompasses both coin tossings, one says that these three possibilities for
the number of heads observed in a tossing are not mutually exclusive,
since, for example, there can be both an observation of 1 head (in one
tossing by one observer) and of 2 heads (in the other tossing by the other
copy of the observer).  Therefore, the three projection operators are not
orthogonal, and the sum of their expectation values can be greater than
unity.

If one had observational access to both coin tossings, one could avoid this
problem by taking a finer-grained set of projection operators, each the
product of the projection operator onto a certain number of heads for a
particular one of the tossings of two coins and of the projection operator
onto a certain number of heads for the other one of the two tossings of two
coins.  Then one would get nine projection operators, one for each ordered
pair of the number of heads for each of the two tossings.  These nine
projection operators are all orthogonal, and their expectation values will
sum to unity if the quantum state gives no other possibilities (e.g.,
possibilities that not both coins are tossed twice or that not all coins
each fall heads or tails).

This all works well in laboratory experiments in which one has
observational access to all the relevant possibilities.  However, in
cosmology in which there may be experiments being made far away by distant
copies of the observer, for which one does not have observational access by
the copy here on earth, one cannot distinguish all of the alternatives
corresponding to a full set of orthogonal projection operators.  For
example, in the coin-tossing experiment in which two tossings occur, one
might only have access to the observation of the number of heads for one
tossing, and one might not even be able to distinguish which tossing one is
observing (e.g., the copies of the observer making the observation might
not have any distinguishable data).  Then one cannot construct from
standard quantum theory projection operators which distinguish the distinct
observations (whether 0, 1, or 2 heads) and which also are orthogonal.  As
given in the example above, if one uses projection operators onto the set
of the three possible distinct observations, they are not orthogonal and
can have expectation values whose sum exceeds unity.

When the probabilistic reasoning of Hartle and Srednicki is cast into the
language of standard quantum theory, it uses the following technique to
avoid the problem of the nonorthogonality of the set of projection
operators for the distinct three observation possibilities:  It uses the
actual observation of the one observer to select the corresponding
projection operator and its complementary projection operator (the identity
operator minus the projection operator onto the actual observed result). 
These two projection operators certainly are orthogonal and sum to the unit
operator, so in a quantum state which is normalized (which we shall always
assume), the sum of the expectation values of these two projection
operators is unity.  However, these two orthogonal projection operators do
not correspond to the results that are observationally distinguishable to
any single copy of the observer.

For example, assume that there are two tossings of two fair coins, but that
the observer of one of the two tossings cannot distinguish which of the two
tossings he or she is observing.  (The two tossings might be observed by
two copies of a locally identical observer, very distantly separated in a
huge spacetime so that the two copies cannot communicate with each other.) 
Suppose that for one of the tossings of two coins, the observer observes a
total of one head.  Hartle and Srednicki make the interpretation \cite{HS},
{\it ``All we know is that there exists at least one such region containing
our data.''}  Therefore, they would calculate the probability for the
existence of one head (out of two coins tossed per tossing) in either or
both of the tossings, which for fair coins would be 3/4.  This would be the
expectation value of the projection operator onto the existence of one head
in either or both of the two tossings of two coins each.  The complementary
probability would be 1/4, the expectation value of the complementary
projection operator onto the nonexistence of exactly one head in either of
the two tossings of two coins each.

However, this complementary projection operator cannot be tested by a
single copy of the observer, since even if it finds that the number of
heads in its tossing is not one, it cannot know whether or not the other
distant tossing gets a result of just one head out of the two coins
tossed.  Therefore, although the one copy of the observer can confirm the
existence of one head (if that is what it observes), it cannot falsify the
existence of one head (no matter what it observes).

On the other hand, if the observer wants a set of projection operators for
which in principle it can confirm any one of them, it could use the three
projection operators onto the existence of 0, 1, and 2 heads respectively. 
However, these are not orthogonal, and for fair coins their expectation
values sum to 13/8, greater than unity.  Therefore, these three expectation
values for the three projection operators whose positive results can be
confirmed by the one observer cannot be interpreted as normalizable
probabilities.

As I see it, this apparent consequence of standard quantum theory and of
similar probabilistic reasoning that has been beautifully deduced by Hartle
and Srednicki \cite{HS} seems to be a {\it reductio ad absurdum} of
standard quantum theory and similar reasoning for cosmologies in which
there are indistinguishable copies of observational situations.  However,
we shall see below that replacing standard quantum theory by Eventual
Quantum Theory can rectify the situation.

\section{General analysis of standard quantum theory}

We can generalize this discussion to the case in which there are $m$
distinguishable possible observations (labeled by a subscript $i$ that runs
from 1 to $m$) in each of $N$ observationally indistinguishable
observational situations.  (That is, which of the $N$ situations is being
observed cannot be distinguished by the identical copies of the observer,
but only the observation outcome itself.)  For notational purposes, suppose
each observational situation is labeled by a superscript index $K$ that
runs from 1 to $N$.  (We might suppose that in principle $K$ can be
determined by some hypothetical super-observer, but not by the ordinary
observer confined to a particular observational situation.)

Now suppose $P_i^K$ denotes the projection operator in the entire quantum
state space onto the $i$th observation in the $K$th situation.  This would
be the tensor product of the local projection operator onto the $i$th
observation in the $K$th local observation situation and of the local
identity operators in all the other local observation situations and in all
other regions of spacetime.  For fixed situation $K$, the different
projection operators $P_i^K$ for different values of $i$ will be assumed to
be orthogonal, $P_i^K P_j^K = 0$ for $i\neq j$, because for a fixed
observational situation (fixed copy of the observer), the different
possible observations are assumed to be mutually exclusive.

However, projection operators for different $K$'s (different observation
situations for different observers) will not be orthogonal, even if their
$i$'s are different:  $P_i^K P_j^L \neq 0$ for $K \neq L$, even if $i \neq
j$.  In fact, if $\langle O\rangle$ denotes the expectation value of the
operator $O$, then although $\langle P_i^K P_j^K \rangle =
\delta_{ij}\langle P_j^K \rangle$ when both projection operators apply to
the same situation $K$, when they apply to different situations $K \neq L$
(here assumed to be in separate local regions, with the quantum state space
a tensor product of the state spaces for each local region), one gets
$\langle P_i^K P_j^L \rangle = \langle P_i^K \rangle \langle P_j^L
\rangle$, the product of the expectation values for the individual
projection operators, which need not be zero.

Because the observer within one observational situation (one copy of the
observer) cannot observe which particular situation he or she is in and
therefore has no access to the index $K$ that is known only to the
hypothetical super-observer, he or she has no justification for using any
particular projection operator $P_i^K$ associated with a particular $K$. 
However, casting the reasoning of Hartle and Srednicki \cite{HS} into
quantum language, one can construct the projection operators $P_i = I -
\prod_K (I - P_i^K)$ onto the existence of the observation $i$ in at least
one of the observational situations, where $I$ is the identity operator for
the full quantum state space, and where $\prod_K$ denotes the product over
all $K$ from 1 to $N$.  If the observer does observe $i$, that would
confirm the truth value of the corresponding projection operator $P_i$, but
its complement, $I - P_i$, cannot be confirmed by any observation
restricted to a single observational situation.

One can take $p_S(i) = \langle P_i \rangle$ to be the probability in
standard quantum theory (denoted by the subscript $S$) that at least one
observation of $i$ occurs, and $p_S(\neg i) = \langle (I - P_i) \rangle = 1
- p_S(i)$ to be the probability that no observation of $i$ occurs. 
However, since for $N \geq 1$ the different $P_i$'s are not orthogonal, the
sum of the $p_S(i)$'s generically will not be unity.  One can follow Hartle
and Srednicki \cite{HS} and say that one has normalizable probabilities
$p_S(i)$ and $p_S(\neg i)$ for any particular $i$, but one can only test
these if one uses the value of $i$ actually observed.  With some
probability the existence of $i$ can be confirmed by an observer within a
single observational situation, but the negation of its existence, $\neg
i$, cannot be confirmed at all.  Because of the asymmetry between the
confirmability of $i$ and the nonconfirmability of $\neg i$, it seems
inappropriate to use $p_S(i)$ as a likelihood in a Bayesian analysis.

We can also see, using an example modeled after that in \cite{BFY07},
quantitatively how $p_S(i)$ can be much larger than the expectation value
of $P_i^K$ for any $K$ and hence can be highly misleading to use as a
likelihood in a Bayesian analysis.  For example, take the case in which
twenty coins are tossed by each of a billion widely separated observers ($N
= 10^9$), and let the observational results be the sequence of twenty heads
and tails ($m = 2^{20} = 1\,048\,576$ possibilities).  Let $i=1$ correspond
to the sequence of all tails (0 heads in the tossing of twenty fair
coins).  If one hypothesizes that all the coins are fair, the expectation
value of $P_1^K$ for any $K$ is $2^{-20} \approx 0.954\times 10^{-6}$, less
than one part in a million.  However, $p_S(1) = \langle P_1 \rangle =
\langle I - \prod_K (I - P_1^K) \rangle = 1 - (1 - 2^{-20})^{10^9} \approx
1 - 6.5\times 10^{-415}$.  If one used $p_S(1)$ as a likelihood in a
Bayesian analysis, one might say that its value, very near unity, would
tend to confirm the hypothesis that the coins are fair, whereas after
getting twenty tails in a row (probability less than one in a million if
there were only one tossing of twenty fair coins), it would seem much more
reasonable to interpret this as evidence against the fair-coin hypothesis.

So how can we modify standard quantum theory, which seems to exemplify the
type of reasoning used by Hartle and Srednicki \cite{HS}, to get more
reasonable results, results in which one can get likelihoods that would not
nearly all tend to unity if there were vastly many indistinguishable
observational situations (identical copies of the observer)?

\section{Eventual Quantum Mechanics}

Let me now propose a version of quantum theory in which the probability of
an observation within a particular observational situation does not depend
on how many such situations there are if the quantum state restricted to
each situation (e.g., its density matrix, after tracing over all other
regions) is the same.  Since the basic elements will be the events
observable within an observational situation, I shall call this class of
quantum theories {\it Eventual Quantum Mechanics}, or EQM.

To motivate what I am aiming for, let me propose that the alternatives for
the observations within an observational situation should obey the
following key principles for the set of alternatives:

(1) {\it Prior Alternatives Principle\/} (PAP):

The set of alternatives to be assigned likelihoods by theories $T_i$ should
be chosen prior to (or independent of) the observation $O_j$ to be used to
test the theories.

(2) {\it Principle of Observational Discrimination\/} (POD):

Each possible complete observation should uniquely distinguish one element
from the set of alternatives.

(3) {\it Normalization Principle\/} (NP):

The sum of the likelihoods each theory assigns to all of the alternatives
in the chosen set should be unity,
\begin{equation}
\sum_j P(O_j|T_i) = 1  . 
\label{norm}
\end{equation}

I am always assuming that the alternatives within any set to be considered
are mutually exclusive and exhaustive (complete).  For example, if the
alternatives are observed data sets within some class, then each
alternative data set must be different, and all data sets within the class
must be included within that set of alternatives.

For pedagogical simplicity, assume initially that the universe does have
$N$ observational situations that are sufficiently indistinguishable that
the observer within each one cannot distinguish which one is his or hers. 
(For example, the distinction might be only in terms of what the
surroundings are at sufficiently great distances that the observer within
the region does not have observational access to this information.) 
Suppose that in each situation there are $m$ distinguishable observations,
say given in the $K$th situation by the $m$ projection operators $P_i^K$
for $i$ running from 1 to $m$ ($P_i^K$ each acting on the entire quantum
state space, but trivially as the identity operator outside the $K$th
observation situation).  Assume they are all orthogonal for each different
$i$ (but the same $K$), $P_i^K P_j^K = \delta_{ij} P_i^K$, and that they
sum to the identity operator $I$ for the entire quantum state space when
summed over $i$, $\sum_i P_i^K = I$ for each fixed $K$.

Now construct the operator $R_i = \sum_K P_i^K$, the sum of the projection
operators over all observational situations $K$ but for the same
observation $i$.  Then define the probability of the observation $i$ in
Eventual Quantum Mechanics as the normalized expectation value of this sum
of projection operators,
\begin{equation}
p_E(i) = \frac{\langle R_i \rangle}{\sum_j \langle R_j \rangle}
= \frac{\langle \sum_K P_i^K \rangle}{\sum_j \langle \sum_K P_j^K \rangle}
= \frac{1}{N} \sum_K \langle P_i^K \rangle.
\label{EQMprob}
\end{equation}

If one has only one observational situation, $N=1$, as has been the usual
implicit assumption in traditional formulations of quantum theory, then of
course $R_i$ is just the projection operator for the observation in that
one situation, and one has the usual probability interpretation of standard
quantum theory.  Thus in that situation, Eventual Quantum Mechanics reduces
to ordinary standard quantum theory.

It is also easy to see that if all the $N$ regions have the same quantum
state (e.g., the same density matrix) and if the $P_i^K$'s are all
essentially the same, except for the specification of which region it is on
which the specific $P_i^K$ acts nontrivially, then $\langle P_i^K \rangle$
is the same for each $K$, and $\langle R_i \rangle$ is just $N$ times this
expectation value.  Therefore, in this case $p_E(i)$ would be the same as
$\langle P_i^K \rangle/\sum_j \langle P_j^K \rangle = \langle P_i^K
\rangle$ for any $K$, the last equality being true because $\sum_j P_j^K =
I$ and $\langle I \rangle = 1$ in a normalized quantum state.  Thus in the
case of $N$ identical regions, Eventual Quantum Mechanics reduces to what
ordinary standard quantum theory would predict for a single one of these
regions.  On the other hand, Eventual Quantum Mechanics does {\it not}
reduce to what standard quantum theory predicts for many such regions, as
has been nicely shown by Hartle and Srednicki \cite{HS}, because the $R_i$
are {\it not} projection operators that are used in standard quantum theory
to give probabilities.

Where Eventual Quantum Mechanics would allow more general predictions than
standard quantum theory applied to a single observational situation would
be in cases in which the different regions (the different observational
situations for different observers) have different density matrices.  Then
the EQM probabilities $p_E(i)$ would be the average of the expectation
values of the projection operators $P_i^K$ over the $N$ regions, an average
probability for the observation $i$ in each of the $N$ regions.

Moreover, one might further generalize Eventual Quantum Mechanics beyond
the last expression of Eq. (\ref{EQMprob}) to allow that the existence of
each region, or the existence of the observer within each region, might
itself have a quantum uncertainty and hence a probability less than unity. 
This could be reflected in the normalization of the effective density
matrix for each region and in the possibility that one might define the
$P_j^K$ more generally so that $\sum_j \langle P_j^K \rangle$ does not
necessarily equal one for each region.  Then the fundamental operator
$R_i$, whose normalized expectation value gives the probability of the
observation $i$, might not be simply a sum of projection operators, but
perhaps a weighted sum of projection operators, where the weights could
effectively give the probabilities of the different regions being realized,
or of the existence of the observer within each of the different regions.

The weights for each region, or for the observer within each region, need
not even be existence probabilities.  For example, they might instead
reflect how long each region lasts, or how long the observer lasts within
each region.

So far as I can see, the main essential feature is that one have a positive
operator $R_i$ for each possible event or observation (or at least an
operator $R_i$ whose expectation value is positive for the actual quantum
state of the cosmos, even if it need not be positive for all possible
quantum states, though it might seems simpler just to require that each
$R_i$ be positive).  It might be easiest to understand the cases in which
each operator $R_i$ is a sum of projection operators, or perhaps a weighted
sum of easily understandable projection operators, but I do not see that
such a requirement would be essential.

\section{Making Eventual Quantum Mechanics More Standard}

One might try to interpret Eventual Quantum Mechanics in a way that appears
more nearly like standard quantum theory.  For example, first consider the
case in which each region, and its observer, definitely exists with unit
quantum probability.  Then although the probability $p_E(i)$ given by Eq.
(\ref{EQMprob}) cannot in general be written as the expectation value of
any natural projection operator for the problem, it could simply be written
as the expectation value of the projection operator $P_i^K$ for any
particular choice of the region $K$ {\it if} the quantum state were
independent of the labeling of the different regions.  That is, if one
replaced the arbitrary density matrix $\rho$ for the tensor product of the
$N$ regions and for whatever else exists outside them by the density matrix
$\bar{\rho}$ that is the average of $\rho$ over all $N!$ permutations of
the regions, then $p_E(i)$ is simply the expectation value of any one
$P_i^K$ (arbitrary $K$) in the averaged state $\bar{\rho}$:
\begin{equation}
p_E(i) = \mathrm{tr}(P_i^K \bar{\rho}).
\label{EQM2prob}
\end{equation}

However, even this conversion of $p_E(i)$ to an expectation value of a
projection operator by changing the state to an averaged state fails to be
true when the quantum probability for the existence of the observer is not
unity for each region.  If in that case one calculates by Eq.
(\ref{EQM2prob}) each $p_E(i)$ and sums them over all possible observations
$i$, the sum will not be normalized to unity but will be the average
probability of the existence of the observer for each region.  To get a
normalized set of probabilities $p_E(i)$ in that case, one should instead
in the averaged quantum state $\bar{\rho}$ take each $p_E(i)$ to be the
conditional probability for the observation $i$ in any one of the regions
(say $K$), given that there is an observer in that region.  If $P_O^K$ is
the projection operator onto the existence of an observer in the region
$K$, then instead of Eq. (\ref{EQM2prob}), one should use the conditional
probability
\begin{equation}
p_E(i) = \frac{\mathrm{tr}(P_O^K P_i^K P_O^K \bar{\rho})}
{\mathrm{tr}(P_O^K \bar{\rho})}.
\label{EQM3prob}
\end{equation}

One might go even further and define the observer-conditioned density matrix
\begin{equation}
\hat{\rho} = \frac{P_O^K \bar{\rho} P_O^K}
{\mathrm{tr}(P_O^K \bar{\rho})},
\label{rhohat}
\end{equation}
so that then the conditional probability $p_E(i)$ can be written as the
expectation value of any one of the projection operators $P_i^K$:
\begin{equation}
p_E(i) = \mathrm{tr}(P_i^K \hat{\rho}).
\label{EQM4prob}
\end{equation}

However, being able to write $p_E(i)$ as the expectation value of a
projection operator in Eventual Quantum Mechanics involves replacing an
arbitrary quantum state $\rho$ with the conditionalized averaged quantum
state $\hat{\rho}$.  If one wanted to stick with the original quantum state
$\rho$, the probability $p_E(i)$ for the observation $i$ (normalized out of
all possible observations) cannot in general be written as the expectation
value of any natural projection operator.

\section{Sensible Quantum Mechanics}

One subclass of Eventual Quantum Mechanics theories are those of Sensible
Quantum Mechanics \cite{SQM1,SQM2,Page-in-Carr,SQM0,SQM3,SQM4} or Mindless
Sensationalism \cite{MS}, in which the alternative events or data sets or
observations are conscious perceptions.  Roughly, each individual conscious
perception is all that a conscious observer is aware of or consciously
experiencing at once, what Lockwood \cite{Lo} calls a ``phenomenal
perspective'' or ``maximal experience'' or ``conscious state,'' and what
Bostrom \cite{Bostrom} calls an observer-moment.  If this conscious
perception is regarded as a observed data set, the data would be the
content of that awareness.  In this set of alternatives, each different
possible conscious perception would be a member, and any two perceptions
with different contents would be different observations.

In the case of a discrete set of conscious perceptions, a particular
Sensible Quantum Mechanics theory assigns a probability to each conscious
perception that is the expectation value of a corresponding positive
`awareness operator.'  There is no requirement that these positive
operators be orthogonal to each other or even that they be proportional to
projection operators (though they might be approximately proportional to
the integral over all of spacetime and over the local Lorentz group of
projection operators in local regions).

\section{Conclusions}

Whether typicality is likely depends on the way likelihoods are
calculated.  The way likelihoods are calculated depends on the set of
alternatives to be assigned likelihoods.  The set of alternatives must be
chosen even before one can do a Bayesian analysis, so one cannot compare
different theories with different sets of alternatives.

Hartle and Srednicki \cite{HS} use a type of probabilistic reasoning that
includes standard quantum theory to select alternatives that in the quantum
case are given by orthogonal projection operators, but their alternatives
are not distinguished by the possible observations in universes large
enough to have many observational situations so similar that they are not
distinguishable to the observers within them.

On the other hand, alternatives obeying the Prior Alternatives Principle
and the Principle of Observational Discrimination have normalized
likelihoods in which typicality is automatically favored in the
likelihoods.  Since this preference comes directly from the likelihoods
normalized over all possible distinguishable observations or data sets, it
is not and need not be introduced ``through a suitable choice of priors''
as Hartle and Srednicki \cite{HS} suggest.  Instead, the prior
probabilities for theories may be chosen to ``favor theories that are
simple, beautiful, precisely formulable mathematically, economical in their
assumptions, comprehensive, unifying, explanatory, accessible to existing
intuition, etc.\ etc.,'' as Hartle and Srednicki propose.

Cosmological theories obeying the Prior Alternatives Principle, the
Principle of Observational Discrimination, and the Normalization Principle,
but apparently not standard quantum theory for a very large universe, can
in principle be tested by observations.  It therefore seems quite
reasonable to adopt these principles.  Eventual Quantum Mechanics and its
subclass of Sensible Quantum Mechanics are frameworks for quantum theories
which do obey these principles and which would make typical observations
more likely.  That is, they enable typicality to be derived as likely.

Typicality by itself does not guarantee that the theory with the highest
posterior probability will make us typical.  However, typicality is
favored in the likelihoods.  One need not impose it separately, but in
discussions in which one does not explicitly invoke the full Bayesian
framework, assuming typicality may be a legitimate shortcut for selecting
between different theories for our observations.  We are unlikely to be
highly atypical.

\section*{Acknowledgments}

I am most grateful to James Hartle and Mark Srednicki for many dozen email
communications.  The hospitality of the George P.\ and Cynthia W.\ Mitchell
Institute for Fundamental Physics of the Physics Department at Texas A\&M
University, and of the Beyond Center for Fundamental Concepts in Science of
Arizona State University, were appreciated for offering me opportunities to
speak in person on this subject with Hartle (and with other persons such as
Gary Gibbons and Seth Lloyd).  The hospitality of Edgar Gunzig and the
Cosmology and General Relativity symposium of the Peyresq Foyer d'Humanisme
in Peyresq, France, enabled me to talk to Brandon Carter and others there
about these ideas.  The hospitality of The Very Early Universe 25 Years On
meeting at the Centre for Mathematical Sciences of the University of
Cambridge gave opportunities for further useful conversations with Hartle
and Alex Vilenkin on this subject.  The hospitality of the University of
California at Santa Barbara permitted conversations in person on the
content of this paper with James Hartle and Mark Srednicki.  Raphael Bousso
and Sean Carroll have also provided helpful email discussions.  More
recently, the hospitality of the Mitchell family at their Cook's Branch
Conservancy enabled me to have many further discussions on this subject
with Hartle and also with Alan Guth and Juan Maldacena.  Finally, the
suggestions of an anonymous referee have helped me clarify my wording. 
This research was supported in part by the Natural Sciences and Engineering
Research Council of Canada.

\end{document}